## *Title:*

Global cross-calibration of Landsat spectral mixture models

## *Author names and affiliations:*

Daniel Sousa[1*] and Christopher Small[1]

[1]Lamont-Doherty Earth Observatory, Columbia University, Palisades, NY 10964 USA

Email: d.sousa@columbia.edu. Tel.: +1 5303044992

## *Abstract:*

Data continuity for the Landsat program relies on accurate cross-calibration among sensors. The Landsat 8 Operational Land Imager (OLI) has been shown to exhibit superior performance to the sensors on Landsats 4-7 with respect to radiometric calibration, signal to noise, and geolocation. However, improvements to the positioning of the spectral response functions on the OLI have resulted in known biases for commonly used spectral indices because the new band responses integrate absorption features differently from previous Landsat sensors. The objective of this analysis is to quantify the impact of these changes on linear spectral mixture models that use imagery collected by different Landsat sensors. The 2013 underflight of Landsat 7 and Landsat 8 provides an opportunity to cross calibrate the spectral mixing spaces of the ETM+ and OLI sensors using near-simultaneous acquisitions of radiance measurements from a wide variety of land cover types worldwide. We use 80,910,343 pairs of OLI and ETM+ spectra to characterize the Landsat 8 OLI spectral mixing space and perform a cross-calibration with Landsat 7 ETM+. This new global collection of Landsat spectra spans a greater spectral diversity than those used in prior studies and the resulting Substrate, Vegetation, and Dark (SVD) spectral endmembers (EMs) supplant prior global Landsat EMs. We find only minor ($-0.01 < \mu < 0.01$) differences between SVD fractions unmixed using sensor-specific endmembers. RMS misfit fractions are also shown to be small (<98% of pixels with <5% root mean square error), in accord with previous studies using standardized global endmembers. Finally, vegetation is used as an example to illustrate the empirical and theoretical relationship between commonly used spectral indices and subpixel fractions. SVD fractions unmixed using global EMs thus provide easily computable, linearly scalable, physically based measures of subpixel land cover which can be compared accurately across the entire Landsat 4-8 archive without introducing any additional cross-sensor corrections.

## 1. Introduction

The Landsat program provides the longest continuous record of satellite imaging of the Earth available to the scientific community (Wulder et al. 2016). One great strength of this record lies in data continuity provided by the generally excellent cross-calibration between the sensors on board the different satellites (Markham and Helder 2012). To extend this continuity into the future, the Operational Land Imager (OLI) onboard Landsat 8 must be intercalibrated with the rest of the archive. Over the 3+ years since launch, the OLI has been shown to exhibit superior performance to previous Landsat sensors with respect to radiometric calibration (Mishra et al. 2016; Morfitt et al. 2015), signal to noise (Knight and Kvaran 2014; Morfitt et al. 2015; Schott et al. 2016), and geolocation (Storey et al. 2014).

One of the applications enabled by such a deep archive of high quality Earth observation data is multitemporal analysis to study long-baseline changes (Vogelmann et al. 2016). However, concern has recently emerged over the direct intermixing of data collected by both the OLI and older TM/ETM+ instruments onboard Landsats 4-7 because of the changes in band placement introduced with Landsat 8 (Holden and Woodcock 2016). Statistical corrections and corresponding transfer functions have been introduced to correct for these differences (Roy et al. 2016). Considerable work has been done to examine the effect of these discrepancies and corrections in the context of spectral indices but, to our knowledge, no attempt has been made to address the implications for multi-sensor or multi-temporal spectral mixture analysis (SMA).

The purpose of this study is to characterize the global Landsat 8 OLI spectral mixing space and cross-calibrate it with the Landsat 4-7 TM/ETM+ spectral mixing space. Previous work has shown the TM and ETM+ sensors to provide globally consistent results for Substrate, Vegetation, and Dark (SVD) subpixel fraction estimates using SMA (Small 2004; Small and Milesi 2013). Extending this cross-calibration to include imagery from the OLI onboard Landsat 8 could thus extend this consistency across the entire 30+ year archive of Landsat 4-8 imagery. In order to develop a cross calibration suitable for multi-sensor SMA, it is necessary to compare spectral mixing spaces for both sensors and identify comparable spectral endmembers that span both spaces. Under ideal circumstances, this would require a spectrally diverse collections of TM/ETM+ and OLI spectra where both sensors image the same targets simultaneously.

Before Landsat 8 was placed into its final orbit, it was maneuvered into underflight configuration below Landsat 7 for one day: March 30 (JD 89) 2013. While the two satellites were positioned in this way, they imaged a diversity of land cover spanning a wide range of spectral reflectance signatures. Each pair of ETM+/OLI images was collected approximately 2-5 minutes apart. The short temporal baseline between image pairs minimizes changes in solar illumination, surface processes and atmospheric effects. The underflight imagery thus provides a rare, nearly ideal opportunity for cross-calibration of the OLI and ETM+ sensors.

In this study, we use 80,910,343 broadband spectra imaged nearly simultaneously by Landsat 7 and Landsat 8 while flown in underflight configuration to address to address the following question: How reliably can subpixel Substrate, Vegetation and Dark (SVD) fractions be used interchangeably between ETM+ and OLI?

We find that the subscenes chosen for this analysis span an even greater range of the Landsat spectral mixing space than previous (Small 2004; Small and Milesi 2013) studies. We suggest that EMs generated for this study can thus effectively replace previous global EMs. While the new Dark (D) EM does not differ substantially from previous EMs, small differences in the Vegetation (V) EM and larger differences in the Substrate (S) EM are apparent. The differences in the Vegetation EM are consistent with the findings of (Holden and Woodcock 2016; Roy et al. 2016) as being a result of band placement. The differences in the Substrate EM are likely due to the wider range of global substrates present in this study than in any previous global study and constitute an improvement upon previous global models.

As a result, we find that subpixel estimates of SVD fractions for Landsat 8 using the old and new EMs display strong linear relations, with estimates of subpixel V fraction essentially unchanged and with easily correctible biases for S and D. When compared with the new EMs, all three SVD fractions scale linearly between the sensors with minimal ($\mu = -0.01$ to $0.01$) bias. RMS misfit to the SVD model for both the old and the new EMs is generally small, with > 98% of all pixels showing < 5% error.

Finally, we use vegetation as an example to show the relationship between commonly used spectral indices and subpixel EM fractions produced by SMA of Landsat 8. We suggest that fractions estimated by SMA from global EMs provide easily computable, linearly scalable, physically based measures of subpixel land cover which can be compared accurately across the entire Landsat 4-8 archive without introducing any additional cross-sensor corrections.

## 2. Background

### *a. Implications of Spectral Band Positioning*

The spectral response function of a sensor quantitatively defines its sensitivity to different wavelengths of light. The radiometric design of the Landsat 8 OLI featured an improvement on the previous TM/ETM+ sensors by modifying its spectral response function to narrow and slightly relocate several of the spectral bands. This has the effect of reducing the impact of common atmospheric absorptions which impede imaging the land surface (Mishra et al. 2016). However, it also has the effect of subtly changing the broadband spectrum imaged by OLI for any object which is not spectrally flat over the wavelengths for which the spectral response function was modified.

Figure 1 shows the effect of the different spectral responses of the OLI and ETM+ sensors. Four sample green vegetation spectra (column 1) are shown, as well as four sample mineral spectra (column 3) from the USGS spectral library. The response functions of the two Landsat sensors are plotted as well to demonstrate the portions of the spectrum over which they are sensitive. The narrowing of the NIR and SWIR 1 bands (black and cyan) are evident, as well as a slight adjustment to the position of the SWIR 2 band. Superimposed on each of these spectra are simulated Landsat 7 and 8 broadband spectra computed by convolving the reflectance spectra with the response functions of the sensors as described above.

Column 2 shows the % difference between the OLI and ETM+ reflectances derived from the laboratory spectra. The essential shape and fundamental characteristics of the spectra are all very similar, but perceptible differences in the spectra are detectible.

While the differences in aggregate are generally <5%, for individual bands the differences can approach 10% in some cases.

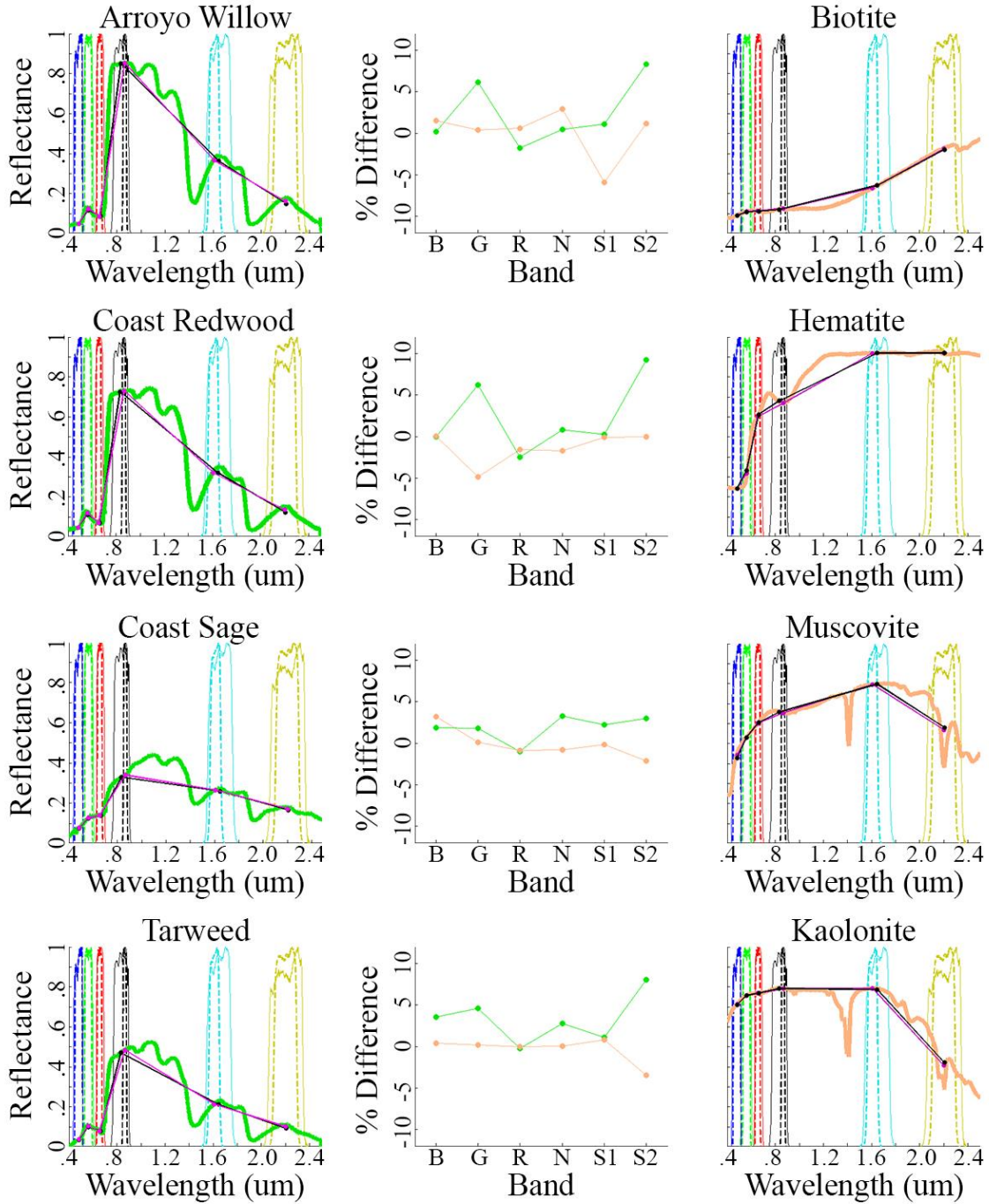

Figure 1. Comparison of Landsat 8 OLI and Landsat 7 ETM+ spectral response functions for vegetation and mineral reflectances. Laboratory spectra from the USGS spectral library for sample vegetation (column 1) and minerals (column 2) convolved with the spectral response functions of OLI and ETM+. The simulated reflectance for each sensor is shown in thick lines (L7 = white, L8 = black). The spectral response functions are generally wider for ETM+ (solid thin lines) than OLI (dashed lines). Percent differences between ETM+ and OLI reflectance (center) depend on both overall albedo and on the depth, width and location of absorptions relative to the response functions. While both sensors record similar spectra, individual band-to-band differences can be >10%.

Figure 1

## *b. Spectral Mixture Models and Linear Spectral Unmixing*

At the scale of the 30 m Landsat pixel, most landscapes are spectrally heterogeneous. As a result, most pixels imaged by Landsat sensors are spectral mixtures of different materials (e.g. soils, vegetation, water, etc) with varying amounts of subpixel shadow. The continuum of aggregate radiance spectra imaged by a sensor forms a spectral mixing space in which each pixel occupies a location determined by the relative abundance of material reflectances imaged in the Ground Instantaneous Field OF View (GIFOV) of the pixel. In situations where multiple scattering among subpixel targets is small compared to single scattering from each subpixel target to the sensor, the aggregate response of the sensor often varies in proportion to the relative abundance of the spectrally distinct materials (Singer and McCord 1979). The topology of the full space of radiance (or equivalently reflectance) spectra reveals the linearity of mixing and the composition of the spectral endmembers and mixtures that bound the space of all other observed spectral mixtures (Boardman 1993). In the case of decameter resolution sensors like those on the Landsat satellites, the combination of spatial and spectral resolution, and positioning of the spectral bands, resolves characteristics of reflectance spectra that distinguish the most spectrally distinct materials commonly found in landscapes. Ice, snow, rock and soil substrates, vegetation, and water each represent a general class of reflectance spectra that are clearly distinguishable with broadband sensors at decameter spatial scales (Small 2004). Of these, the aggregate broadband reflectances of most landscapes can be represented accurately as linear mixtures of substrate (S), vegetation (V) and dark (D) endmembers. The dark endmember corresponds to either absorptive, transmissive or non-illuminated surfaces and typically represents either shadow or water. As a result, linear combinations of these three spectral endmembers can represent the aggregate reflectance of a very wide range of landscapes at meter to decameter scales (Small and Milesi 2013). By identifying the SVD endmember spectra that bound the spectral mixing space, it is possible to use these endmembers together with a linear spectral mixture model to project the 6D feature space of the Landsat sensors onto a simpler 3D mixing space bounded by spectrally and functionally distinct components of a wide range of landscapes (Adams et al. 1986). Inverting a simple three endmember linear spectral mixture model using the SVD endmembers yields estimates of areal abundance of each endmember for each pixel in an image. Using standardized spectral endmembers that span the global mixing space of spectra allows for intercomparison of fraction estimates derived from different sensors across space and time. Standardized spectral endmembers confer all of the benefits of spectral indices, with the added benefit of using all of the spectral information available while simultaneously representing multiple spectral contributions to the mixed pixel.

## 3. *Data & Methods*

All data used in this study were acquired from the USGS Earth Resources Observation and Science Center at http://glovis.usgs.gov/. Landsat 8 data were acquired from the "Landsat 8 OLI Pre-WRS 2" collection. Data were calibrated to exoatmospheric reflectance (Chander and Markham 2003) using the standard ENVI calibration tool. A spectrally diverse set of 100 30 x 30 km subscenes was selected from the spatial overlap between the Landsat 7 and 8 acquisitions. Nearly all of the subscenes were cloud-free, although some subscenes which contained land cover with unusually diverse spectral properties were included even if minor cloud contamination was present. Both Landsat 7 and 8 analyses were performed only on pixels unaffected by the SLC-off gaps. Principal components analysis and linear spectral unmixing were performed using the standard ENVI routines. All unmixing was performed with unit sum constraints with weight = 1.

## *4. Analysis*

Figure 2 shows the locations of the 30 Landsat 7 and 8 scene pairs used in this analysis. All scene pairs were collected in underflight configuration. The time difference between Landsat 7 and 8 overpasses was < 6 minutes for every scene pair. The scenes span a remarkable geographic diversity of land cover, given the short time in which they were collected. Five continents are represented. Although several images were acquired over mainland Europe (Path 198), unfortunately all except the one covering Ibiza, Spain were too cloudy for the purposes of this analysis.

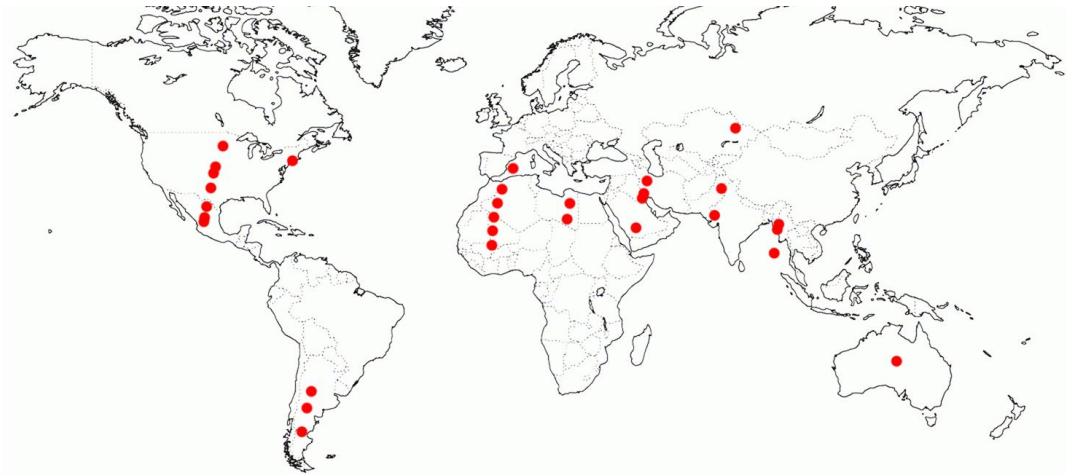

Figure 2. Locations of 30 near-simultaneous Landsat 7/8 scene pairs from which the 100 subscenes for this analysis were chosen. For every scene pair, Landsat 7 and Landsat 8 overpass times were within 6 minutes of each other. All scenes were imaged while Landsat 8 was performing its pre-WRS2 underflight of Landsat 7 on March 30 (JD 89), 2013.

Figure 2

From these 30 image pairs, 100 subscenes were chosen on the basis of spectral diversity (Figure 3). Subscenes are shown both with a common linear stretch (TOA reflectance = 0 to 0.7) and subscene specific 2% linear stretches in an attempt to show the spectral diversity and complexity included in this sample. Shallow and deep water are each represented in both coastal and inland water bodies. Natural and managed vegetation are both present over a wide range of climate zones and soil types. Geologic diversity includes both mafic and felsic bedrock, quaternary alluvium, and sand dunes with variable grain size and lithology. One large evaporite pan near Kuwala, India was included to demonstrate the performance of spectrally complex minerals in the global SVD model. Despite several cloud-free acquisitions at high northern latitudes, snow and ice was minimized due to its minor areal coverage within the terrestrial ecoregions of the world (Olson et al. 2001) and the fact that a larger sample would be required to accurately represent its true spectral diversity. When pixels in the SLC-off gaps of Landsat 7 are removed, a total of 80,910,344 coregistered ETM+ and OLI spectra remain.

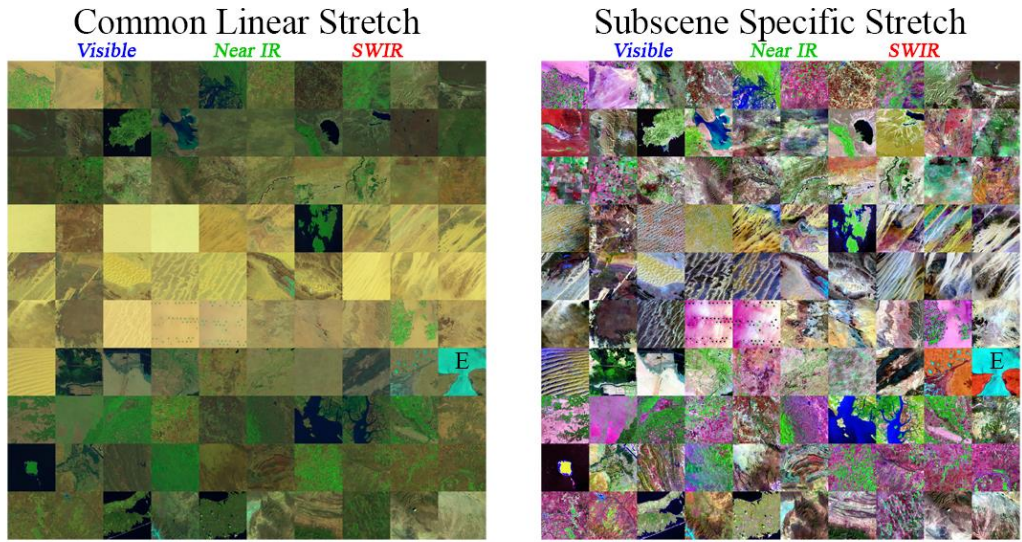

Figure 3. Comparison of 100 OLI subscenes chosen from the near-simultaneous Landsat 7 and Landsat 8 acquisitions from Figure 2. Each 30x30 km subscene is shown with both a common linear stretch (reflectance = 0 to 0.7) as well as with subscene-specific 2% linear stretches to illustrate the spectral diversity of the scenes chosen. The subscenes sample a range of evergreen and deciduous natural vegetation, agriculture, and standing water (both deep and shallow), as well as lithologically variable soil, sediment, and rock substrates. With the exception of the evaporite pan (labeled E) in western India, all subscenes are composed of varying mixtures of rock and soil substrates, vegetation, water and shadow.

Figure 3

Principal Component (PC) analysis was then performed independently on both the Landsat 7 and Landsat 8 subscene mosaics. Landsat 8 Coastal/Aerosol and Cirrus bands were not included in the analysis in order to facilitate a direct comparison between the sensors. The resulting Landsat 8 spectral mixing space with corresponding single pixel EMs is shown in Figure 4. The Landsat 7 mixing space is not shown, as it is visually indistinguishable from the Landsat 8 space. As found in previous work, the space is characterized by sharp apexes corresponding to Vegetation and Dark EMs, but substantially more complexity for the Substrate EM. This complexity reflects the diverse range of rocks and soils spanning the plane of substrates. Sharp edges connecting (D,V) and (D,S) EMs (clearly visible in the projection showing PC 1 and PC 3) indicates binary linear mixing. Concavity on the edge connecting (S,V) suggests that Substrate and Vegetation rarely trade off completely without any subpixel shadow. The elongate cluster of pixels spectrally distinct from the global mixing space corresponds to the Evaporite pan (E) in India. The inclusion of these evaporites allows an opportunity to illustrate the behavior of the model to materials which are not linear combinations of substrate, vegetation, or dark targets in broadband visible-IR spectra.

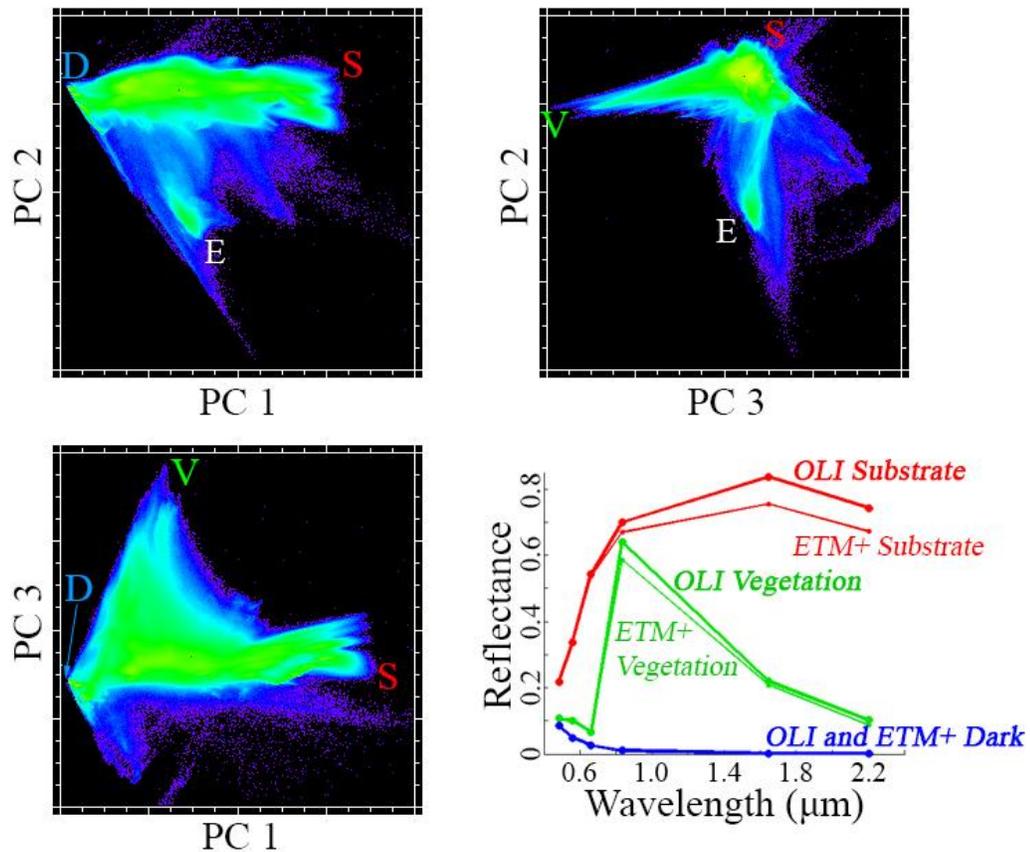

Figure 4. The Landsat 8 OLI spectral mixing space derived from 80,910,343 broadband spectra. The Landsat 7 ETM+ mixing space (not shown) of the near-simultaneous Landsat 7 acquisitions is visually indistinguishable. Endmember spectra (lower right) selected from the apexes of the scatterplot correspond to the same geographic locations so represent the same materials - within the uncertainty in the coregistration of each OLI/ETM+ image pair. The prominent cluster with distinct PC 2 values (E) corresponds to an evaporite pan near Kuwala, India.

Figure 4

Substrate (red), Vegetation (green) and Dark (blue) global EM spectra are shown in Figure 4. The differences between the ETM+ and OLI EM spectra are a result of the changes in spectral response functions between the sensors. These pairs of spectra represent identical geographical locations imaged at nearly the same time. The Substrate EM corresponds to a field of sand dunes in the Libyan Sahara (p184r044), the Vegetation EM corresponds to a homogenous agricultural field in central Texas (p029r038), and the Dark EM corresponds to deep water off the Atlantic coast of Long Island, New York (p013r032). While the dark EM is nearly identical for the two sensors, the Landsat 8 substrate and vegetation EMs are brighter than the Landsat 7 EMs in all IR wavelengths, most prominently in the NIR.

As expected, the geometry of the mixing space shown here, as well as the spectra of the resulting Vegetation and Dark EMs are similar to those found by previous studies (RMS differences with (Small and Milesi 2013) of 0.02 and 0.00 for V and D, respectively). However, the Substrate EM is substantially brighter across all wavelengths than found previously (RMS differences with (Small and Milesi 2013) of 0.14 for the new OLI EM and 0.10 for the new ETM+ EM). The plane of substrates found in this study is inclusive of the spectral range found by prior studies, but also contains substantially greater variability in bright sands. This extension of the plane of substrates is likely a result of the range of diversity of sands and soils included in this analysis. The newly identified substrates represent an improvement over previous models as they are more general and inclusive of the range of landscapes present on the surface of the Earth.

The newly identified global EMs were used to unmix the collection of OLI and ETM+ underflight spectra. Figure 5 shows the comparison of SVD fraction estimates from Landsat 8 OLI spectra as unmixed using the old (Small and Milesi 2013) global EMs and the new underflight OLI EMs. As expected given the new, more reflective substrate EM, substrate fractions are substantially lower and dark fractions are substantially higher with the new EMs than with the old. Note that the x-axes of the Substrate and Dark plots are truncated at upper bounds of 1.2 and lower bounds of -0.2, respectively. A substantial number of pixels have substrate fractions as high as 1.4 and dark fractions as low as -0.4 when unmixed with the old EMs. The new EMs more effectively span the global mixing space and result in the physically plausible bounds of 1.0 and 0.0 for these fractions.

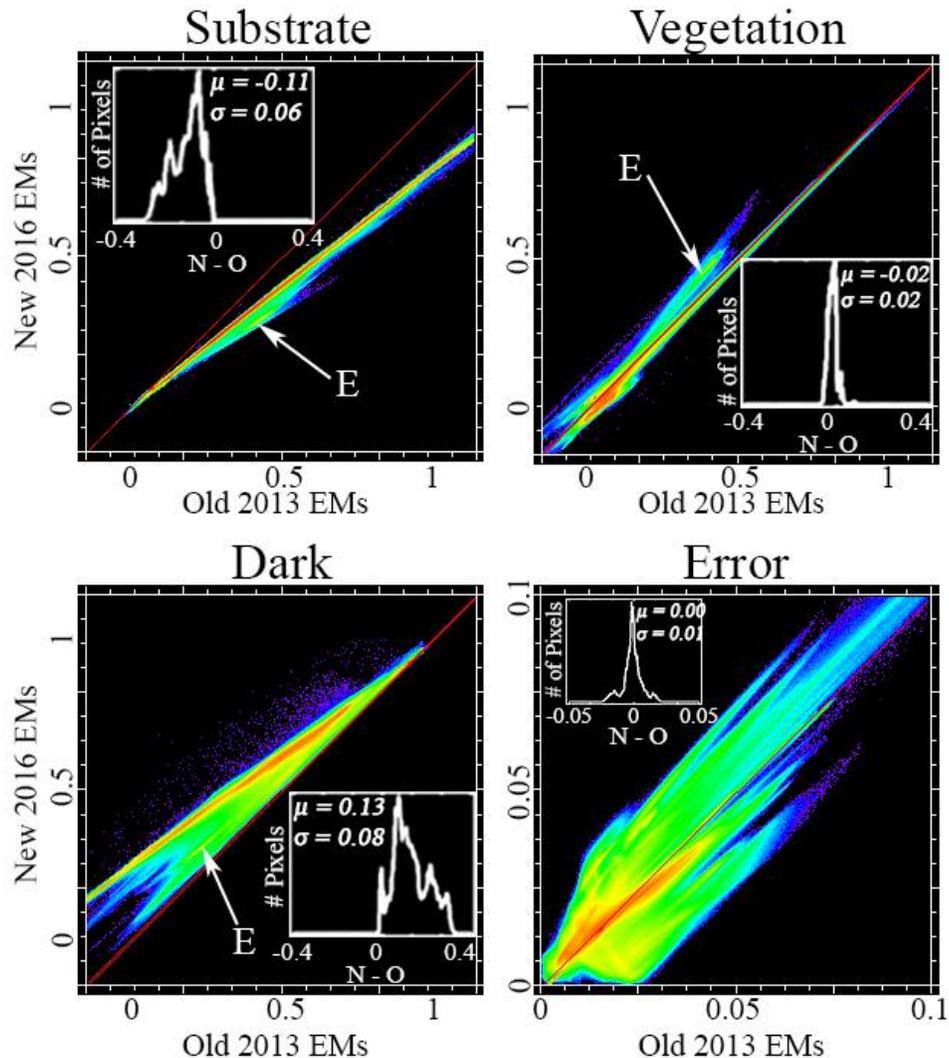

Figure 5. SVD fraction intercomparison for unmixing of 80,910,343 Landsat 8 spectra using both the previous (Small & Milesi 2013) global EMs and the new 2016 EMs. Landsat 8 OLI fractions unmixed with the new OLI EMs and with the old 2013 TM/ETM+ EMs are strongly linear - even though the EMs were selected from independent global collections of spectra. Fractions estimated with the old 2013 EMs show a clear bias toward higher substrate fractions ($\mu = -0.11$) and lower dark fractions ($\mu = 0.13$) than those estimated with the new 2016 EMs. Vegetation fraction shows a small bias.($\mu = -0.02$). Error fractions are slightly lower for the new underflight EMs than for the previous global EMs, but >98% of all pixels have error <5% for both models. The cluster of pixels distinctly plotting off the linear relations for S, V, and D fractions corresponds to evaporites (E) which are not well represented by either simple 3 EM model. Histogram insets show fraction difference (New - Old) between the two models.

Figure 5

The vegetation fractions plot close to the 1:1 line, indicating that vegetation estimates are essentially unchanged between the sets of EMs. RMS error estimates are essentially unchanged between the two sets of EMs, with > 98% of all pixels showing error < 5%. The evaporites plot distinctly off the 1:1 line for all fractions, showing reduced S, increased V, and reduced D fractions relative to the rest of the global space. These values are clearly erroneous and reflect the inability of the SVD model to represent evaporites. The evaporite EM is not included in the SVD model because evaporites represent a small fraction of Earth's surface and lie outside the primary SVD hull that represents most landscapes. However, the quasi-linear binary mixing trend between the evaporite and dark EMs suggests that a linear mixture model might be useful for mapping variations in moisture content of evaporites. We do not include an evaporite EM here because our single acquisition is not necessarily representative of the true diversity of evaporites and range of moisture contents. We omit ice and snow EMs for the same reason.

Figure 6 shows the cross comparison between Landsat 8 underflight fractions unmixed using the new OLI global EMs (thick lines from Figure 4) and Landsat 7 underflight fractions unmixed using the corresponding new global ETM+ EMs (thin lines from Figure 4). Biases for all fractions are small (-0.01 < $\mu$ < 0.01) and all fractions cluster tightly around the 1:1 line ($\sigma$ = 0.03 for all fractions and $\sigma$ = 0.00 for error). The small number of pixels plotting substantially off the 1:1 line can generally be visually identified as either: 1) atmospheric effects which changed over the 1-6 minutes between satellite overpasses or 2) land cover types poorly fit by the global SVD model such as snow/ice or shallow/turbid water. The evaporite cluster remains clearly distinct as a reminder of the limits of the model. Some of the dispersion about the 1:1 line may also be attributed to spatial misregistration between Landsat 7 and 8, although visual comparison suggests excellent coregistration in most cases. This suggests that subpixel displacements between the Landsat 7 and Landsat 8 acquisitions may introduce fraction differences of several percent in some cases, although the majority of pixels agree to well within 3%. The linearity, lack of bias, and tight clustering of these scatterplots suggest TM/ETM+ and OLI imagery can be safely used interchangeably when unmixed using these global EMs.

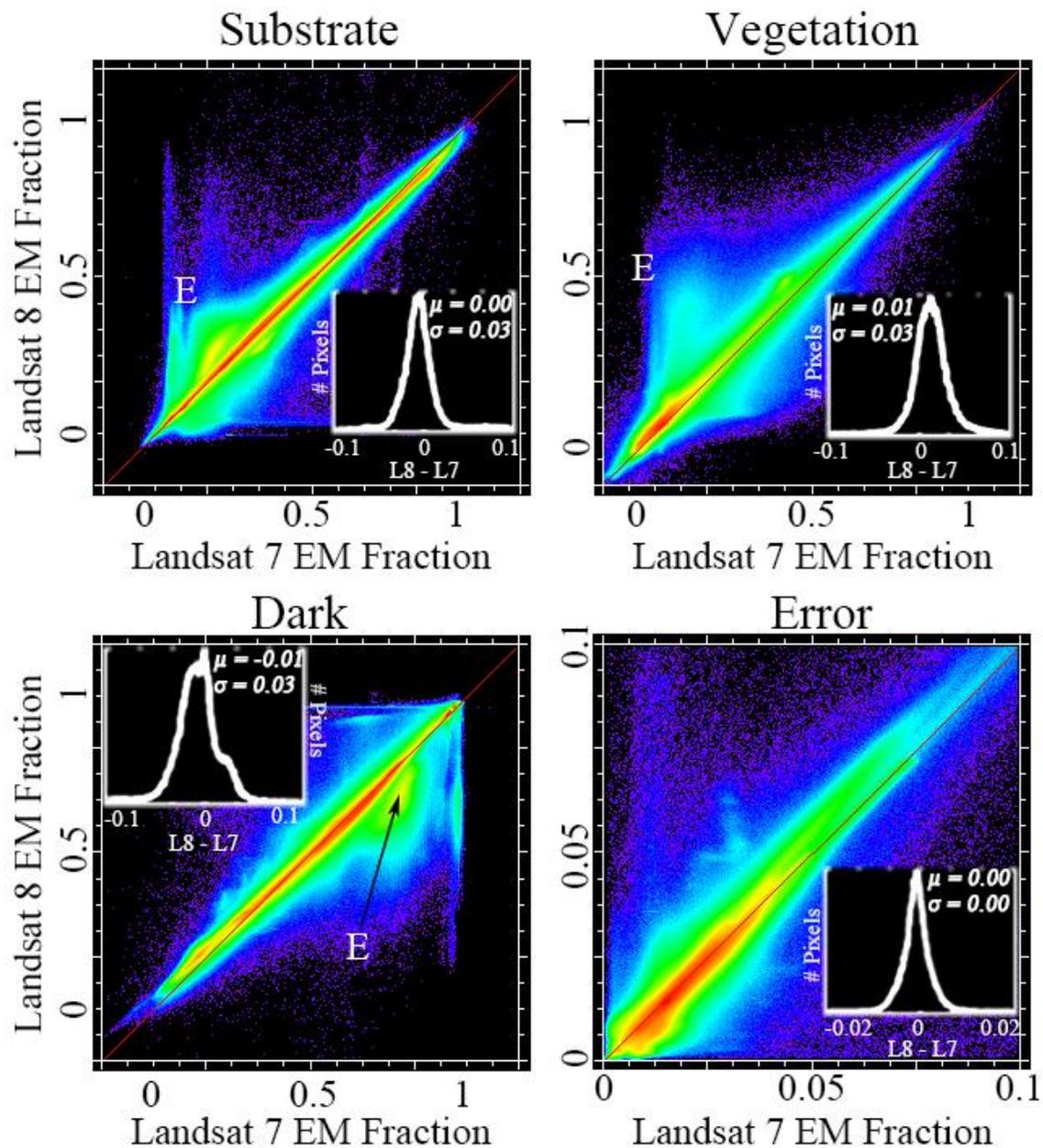

Figure 6. Intercomparison of SVD fractions derived from 80,910,343 near-simultaneous ETM+ and OLI spectra using the new underflight ETM+ and OLI EMs. All fractions (including error) across sensor with minimal bias (μ ≤ 1%) off the 1:1 line. Scatter about the 1:1 line corresponds to either pixels with changing atmosphere in the 1-6 minutes between satellite overpasses or subpixel displacements between images. Evaporites (E) are not well represented by the SVD model so they also plot off axis. Inset histograms show fraction difference distributions. All three SVD fractions show >95% of all pixels with differences in the range +/- 5%. Error differences show 98% of all pixels have < 1% for the SVD model.

Figure 6

As discussed by (Holden and Woodcock 2016), differences in the OLI and ETM+ spectral responses have implications for comparability of spectral indices. We compare three commonly used vegetation indices with vegetation fraction estimates for the diversity of Landsat 8 OLI spectra in the underflight collection. Figure 7 shows the relation between subpixel vegetation fraction (Fv) as estimated with the new global SVD EMs and three commonly used vegetation indices: Normalized Difference Vegetation Index (NDVI, (Rouse et al. 1973)), Enhanced Vegetation Index (EVI, (Huete et al. 2002)), and Soil Adjusted Vegetation Index (SAVI, (Huete 1988)).

The equation used for NDVI is:

$$\frac{NIR - V_r}{NIR + V_r}$$

The equation used for EVI is:

$$2.5 * \frac{NIR - V_r}{NIR + 6 * V_r - 7.5 * V_b + 1}$$

The equation used for SAVI is:

$$1.5 * \frac{NIR - V_r}{NIR + V_r + 0.5}$$

The relationship between SAVI and Fv is relatively linear for most pixels with Fv > 0.2, although a substantial bias is present and variance is wide at low values. The relationship between EVI and Fv is also linear, although with considerable variability and positive offset from the 1:1 line. The relationship between NDVI and Fv is substantially more complex and shows the well-known saturation effect at high vegetation fractions.

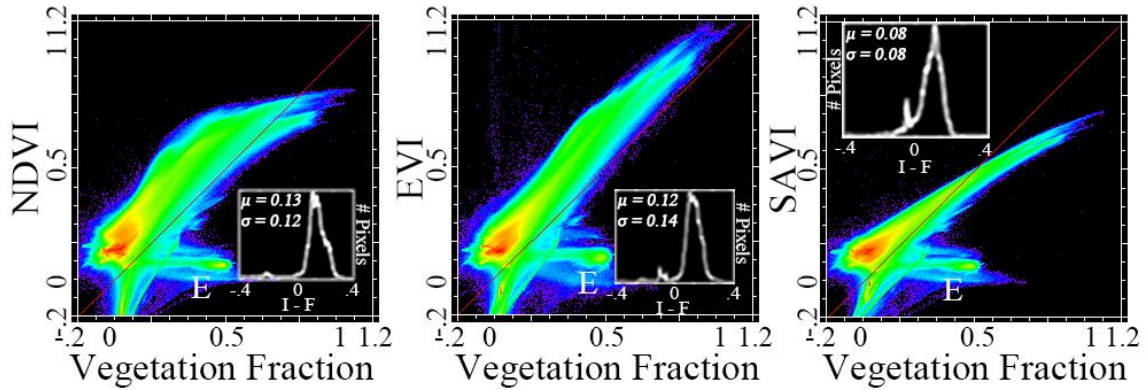

Figure 7. Vegetation index intercomparison. Top row: performance of NDVI, EVI, and SAVI relative to vegetation fraction, as measured by Landsat 8. EVI and SAVI could potentially be corrected to vegetation fration over a portion of their range with linear transformations. NDVI is substantially more complex. Bottom row: NDVI, EVI, and SAVI as measured by Landsat 7 and 8 underflight. All three indices are generally linear as imaged by the two sensors, but all show deviation off the 1:1 line indicating generally higher estimates from Landsat 8 than from Landsat 7. Inset histograms with mean and standard deviation shown as in previous figures. The evaporites (E) remain as a prominent cluster not well represented by any vegetation metric.

Figure 7

## 5. Discussion

The complex relationships of the vegetation indices shown in Figure 7 may not be intuitive given their mathematical simplicity. This complexity is not a function of the geographic limitation of the study or of the limitations of SMA. Instead, the complexity can be shown to have a simple theoretical explanation.

To illustrate the basis for the complexity of these relations, consider a hypothetical 30 x 30 m Landsat pixel filled with some amount of green vegetation and some amount of exposed soil. Based on the solar geometry illuminating the pixel, there will be some variable amount of area (viewed from directly above) of subpixel shadow cast by the roughness of the soil and the height and geometry of the vegetation. Areas in deep shadow are illuminated only by diffuse scattering with a spectrum dominated by Rayleigh scattering in the atmospheric column between the ground and sensor – as illustrated by the Dark EM. Between deep shadow and illuminated substrate and vegetation is a continuous triangular plane of spectral mixtures. This plane includes 100% illuminated vegetation with no soil or shadow, 100% illuminated soil with no vegetation or shadow, and 100% deep shadow (Rayleigh scatter only) – as well as all combinations thereof. The sensor essentially integrates these continuous endmember spectra as a linear sum into a single 6-element broadband spectrum. We use the OLI spectral EMs from Figure 4, with the linear spectral mixture model to simulate all possible mixtures of substrate, vegetation and shadow, then compute vegetation indices (NDVI and EVI) for each simulated mixed pixel.

Figure 8 shows the results of this simulation for every possible mixture of vegetation, soil and shadow in 1% increments, resulting in 5050 simulated Landsat spectra. This simulation is run for 3 different levels of atmospheric "noise" (in the form of adding a random Rayleigh scattering spectrum as the dark EM) as well as 3 different background soils (produced by varying the brightness of the soil spectrum as the substrate EM). NDVI and EVI are then computed for all of these simulated pixels, and $F_v$ is estimated by unmixing using SMA.

As expected, inversion of the linear SVD model yields accurate results for $F_v$, with minimal bias and scatter (in all cases $\mu < 0.5\%$ and maximum error of any pixel $< 2.5\%$), with nearly uniform dispersion across the full range of values. Because of the high degree of linearity between $F_v$ and input area of vegetation, and for ease of comparison to Figure 7, we plot the indices against $F_v$.

The behavior of the vegetation indices is complex. Varying the amplitude of atmospheric noise or the spectrum of the soil substrate can substantially alter the bias and curvature of the indices. Over a wide range of soils, EVI exhibits substantial linearity with Fv, although it consistently plots above the 1:1 line for all but the brightest soil. EVI is also shown to deviate more strongly from linearity with more severe atmosphere, especially at high vegetation fractions. NDVI demonstrates its well-known saturation at high values and greatly variable nonlinear dependency on the soil spectrum.

This range of outcomes for spectral indices with small variations in atmospheric and soil parameters is a result of the functional form of the equations used in the computation of the indices. NDVI is a simple ratio of the sum and difference of 2 bands. EVI introduces a third band, and exhibits substantially enhanced stability over a range of conditions. Fv uses the information content of all 6 bands in the spectrum and explicitly accounts for the contributions of both soil and shadow. This results in enhanced theoretical stability of Fv over indices based on only 2 or 3 bands – stability which also applies to any systematic perturbations which may be introduced by the changes in spectral response between ETM+ and OLI.

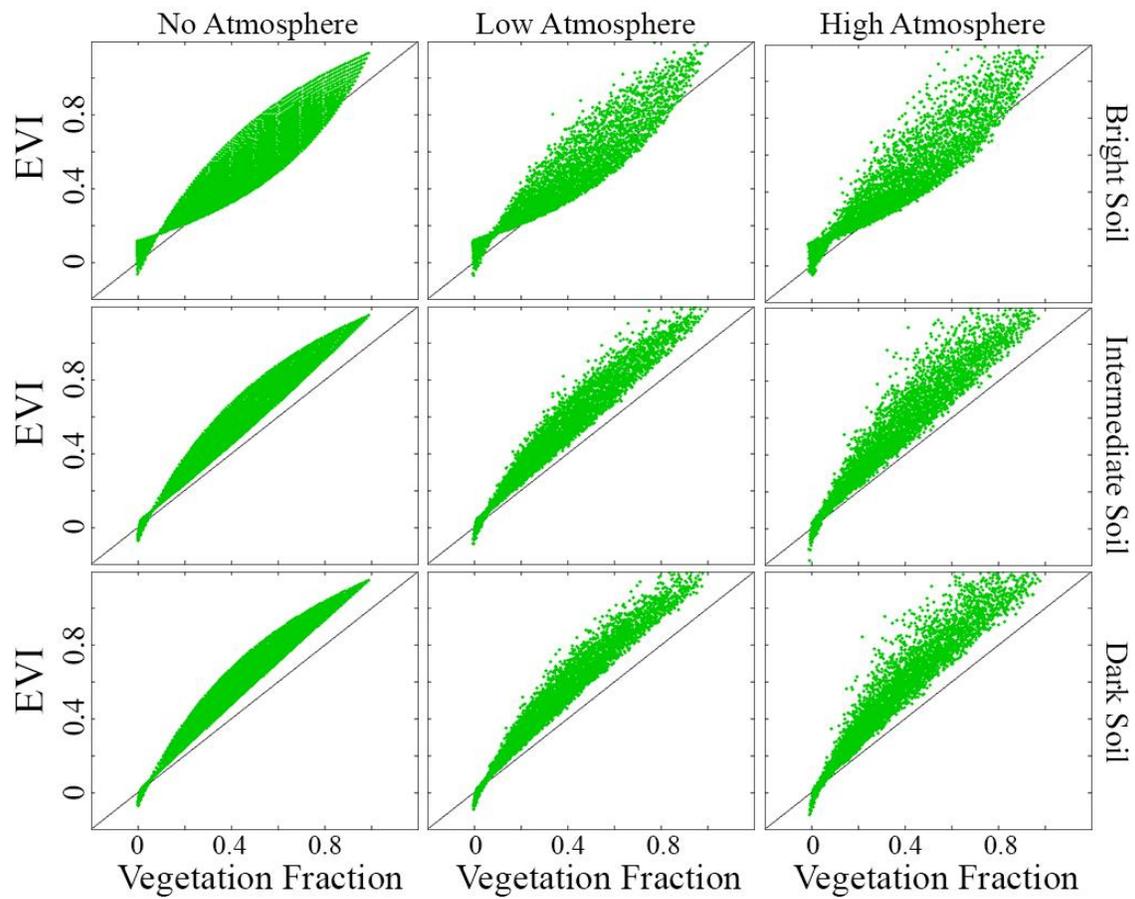

Figure 8a. Calculation of EVI for theoretical pixels contaning every possible integer combination of subpixel soil, vegetation, and shadow. EVI exhibits more linearity over a wider range than NDVI. High values of EVI show sensitivity to atmospheric perturbations.

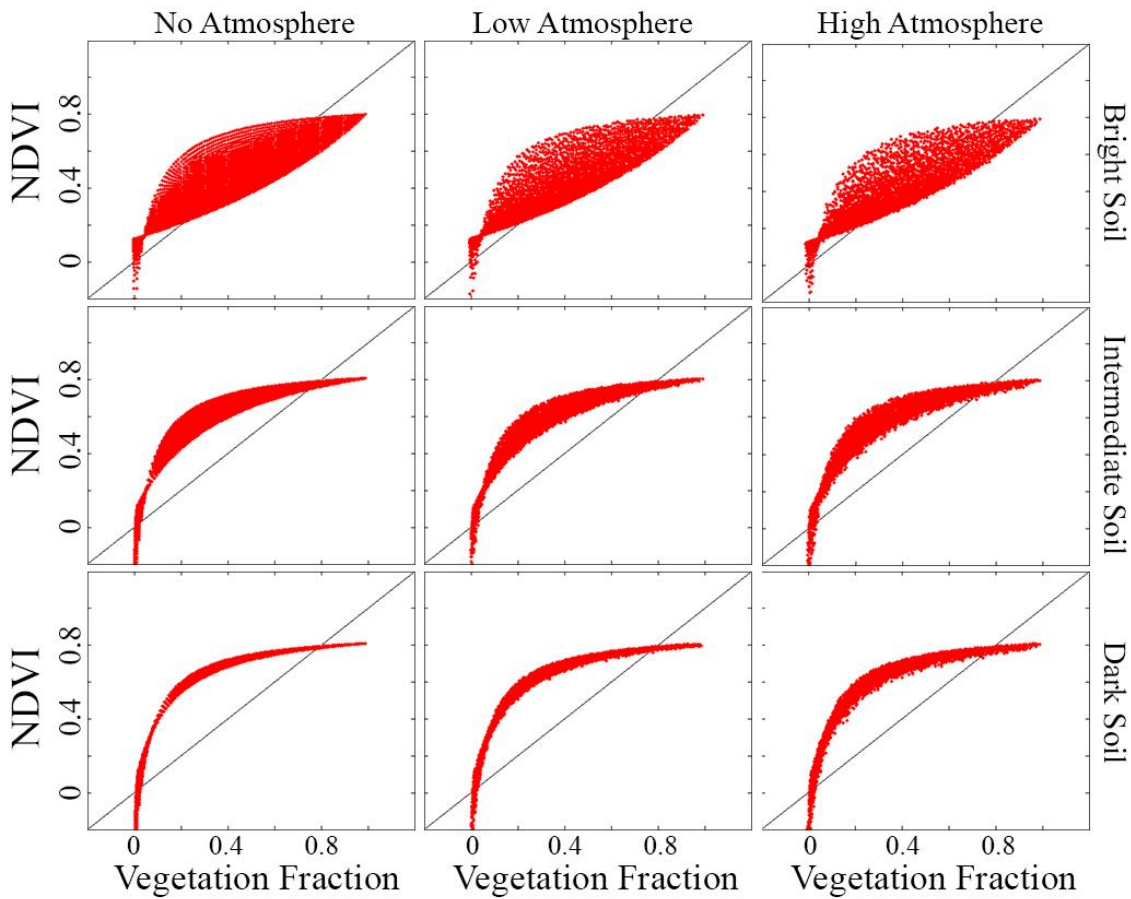

Figure 8b. Calculation of NDVI for theoretical pixels contaning every possible integer combination of subpixel soil, vegetation, and shadow. Slight variations in the amount of atmospheric perturbation (simulated as Rayleigh scatter times a small random number) and brightness of the soil substrate can yield substantial differences in the oucome of the index.

Figure 8

## 6. *Conclusions*

Subpixel EM fractions for Landsats 7 and 8 imaged in underflight configuration over a wide range of land cover show considerable agreement and can be well-characterized by the simple 1:1 relation with minimal bias or scatter. RMS misfit for both sensors using these new models remains < 5% for > 98% of the pixels, as good or better than the previous EMs. It is also notable that no atmospheric correction was attempted for this study (beyond the selection of subscenes which appeared to be cloud-free). This agreement is testament to the work done by those at NASA and the USGS responsible for the design and implementation of the radiometric cross-calibration of these sensors.

The results of the EM fraction comparison suggest that the differences in bandpasses between the two sensors can effectively be taken into account by the use of new EMs based on the near-simultaneous imaging of the same geographical locations by the two sensors – with no additional radiometric cross-calibration. In addition, these EMs now more fully span the global mixing space than previous EMs due to the inclusion of additional bright sands which extend the plane of substrates beyond previous studies. We suggest that these new global EMs supplant the EMs from previous studies. These EMs are freely available online at: [www.LDEO.columbia.edu/~small/GlobalLandsat/](www.LDEO.columbia.edu/~small/GlobalLandsat/)

However, the behavior of spectral indices, as already noted by others, is substantially more complex and may require cross-calibration beyond direct download of L1T imagery from the USGS archive if such indices are to be used operationally to compare TM/ETM+ and OLI imagery, as discussed by (Holden and Woodcock 2016) and (Roy et al. 2016).

## 7. Acknowledgements

The authors thank those responsible for the free availability of the Landsat imagery used in this study, as well as those responsible for the extensive preprocessing of that imagery to facilitate its use. Work done by D. Sousa was conducted with Government support under FA9550-11-C-0028 and awarded by the Department of Defense, Air Force Office of Scientific Research, National Defense Science and Engineering Graduate (NDSEG) Fellowship, 32 CFR 168a. CS was funded by the NASA MultiSource Land Imaging Program (grant NNX15AT65G). D. Sousa thanks M.B. Sousa for helpful and clarifying conversations.